\documentclass[prb,twocolumn,showpacs,preprintnumbers,amsmath,amssymb]{revtex4}
\usepackage{graphicx}
\usepackage{dcolumn}
\begin{document}

\title{ Synchronous sublattice  algorithm for parallel kinetic Monte Carlo}

\author{Yunsic Shim}
\email{yshim@physics.utoledo.edu}
\author{Jacques G. Amar}
\email{jamar@physics.utoledo.edu}
\affiliation{Department of Physics \& Astronomy \\
University of Toledo,  Toledo, OH 43606}

\date{\today}

\begin{abstract}
The standard kinetic Monte Carlo algorithm is an extremely efficient   method to carry out  serial simulations of dynamical processes such as thin-film growth.  However, in some cases it is necessary  to study systems over extended time and length scales, and therefore a parallel algorithm is desired.  Here we describe an efficient, semi-rigorous synchronous sublattice  algorithm for parallel kinetic Monte Carlo simulations.  The  accuracy and parallel efficiency  are studied as a function of  diffusion rate, processor size, and number of processors for  a variety of simple  models of epitaxial growth.  
The effects of fluctuations on the parallel efficiency are also studied. Since only  local communications  are required, linear scaling behavior is observed, e.g. the parallel efficiency is independent of the number of processors  for fixed processor size.  
\end{abstract}
\pacs{81.15.Aa, 05.10.Ln, 05.10.-a, 89.20.Ff}

\maketitle
\section {Introduction}

Kinetic Monte Carlo (KMC) is an extremely efficient method \cite{bkl, voter, maksym,fichthorn, 
blue}  to carry out dynamical simulations of stochastic and/or thermally activated processes when the relevant activated 
atomic-scale processes are known.   KMC simulations have been successfully  used to model a   variety of dynamical processes ranging from catalysis to thin-film growth.   The basic principle of kinetic Monte Carlo is that in order to efficiently simulate a dynamical system with a variety of different rates or processes, at each step in the simulation one picks the next process to occur with a probability proportional to the rate for that process.  The time of the next event is determined by  the total overall rate for all processes to occur, and  after each event the rates for all processes are updated as necessary. 

 In contrast to Metropolis Monte Carlo, \cite{Metropolis} in which each Monte Carlo  step corresponds to a  configuration-independent  time interval and  each event is selected randomly but only accepted  with a configuration-dependent   probability, in kinetic Monte Carlo both the selected event and the time interval between events are configuration-dependent while the acceptance probability  is fixed (all attempts are accepted).   In the context of traditional equilibrium Monte Carlo simulations this is sometimes referred to as the n-fold way. \cite{bkl}  Although KMC requires additional book-keeping to keep track of the rates (probabilities) for all possible events,   the  KMC  algorithm is typically significantly  more efficient than the Metropolis algorithm since no selected moves are rejected.  In particular, for problems such as thin-film growth in which the possible rates or probabilities for events can vary by several orders of magnitude,   the kinetic Monte Carlo algorithm can be orders of magnitude more efficient than Metropolis Monte Carlo.

The standard  KMC algorithm is a  serial algorithm since    only one event can occur at each step. 
However, for some problems one needs to simulate  larger length   and time-scales than can be simulated using a serial algorithm.  For these problems it would be desirable to develop efficient parallel kinetic Monte Carlo  algorithms so that many processors can be used simultaneously  in order to carry out realistic computations over extended time- and length-scales.

Recently there has  been a great deal of  work on asynchronous parallel algorithms for Metropolis Monte Carlo using  domain decomposition. In particular, because the attempt time in Metropolis Monte Carlo is independent of system configuration,  an asynchronous ``conservative" algorithm may be used.  \cite{chandy, Lubachevsky1, Korniss1, Korniss2, Korniss3}  In such an algorithm all processors    whose next attempt time is less than their neighbor's next attempt times are allowed to proceed.  Unfortunately    such a ``conservative" algorithm does not work  for kinetic Monte Carlo since in KMC  the event-time depends on the system configuration.   In particular, since fast events may ``propagate" across processors, 
the time for an event already executed by a processor may change due to earlier events in nearby processors, thus leading to an incorrect evolution.  As a result, the development of efficient parallel algorithms for kinetic Monte Carlo simulations remains a  challenging problem.

A more efficient version of the conservative asynchronous algorithm for Metropolis Monte Carlo has been developed by Lubachevsky\cite{Lubachevsky1} in the context of Ising simulations and has been implemented by Korniss et al.\cite{Kornissjcp}  In this approach, ``n-fold way" \cite{bkl} simulations are carried out in the interior of each processor, while  Metropolis simulations are carried out  at the boundary.  At each step, either an interior move or a boundary move is selected with the appropriate probability. While all ``n-fold way" interior moves are immediately accepted,
all Metropolis attempts must wait until the neighboring processor's next attempt time is later before being either accepted or rejected. Since such an algorithm is equivalent to the conservative Metropolis Monte Carlo algorithm described above, it  is 
 generally scalable,\cite{Korniss1, Korniss2, Korniss3}  and  has    been found to be relatively efficient in the context of kinetic  Ising model simulations in the metastable regime. \cite{Kornissjcp, Kornisse1,Kornisse2}

 Such an approach can be generalized\cite{yshimnovotny} in order to carry out parallel KMC simulations by mapping all KMC moves to a Metropolis move with an acceptance probability given by the rate for that event divided by the fastest possible rate in the KMC simulation. However, because of the possibility of  significant rejection of boundary events, the parallel efficiency  may be very low for problems with a wide range of rates for different processes.  For example, we have recently\cite{yshimnovotny}  used  such a mapping to carry out parallel KMC simulations  of a simple 2D solid-on-solid ``fractal" model of submonolayer growth with a moderate monomer ratio $D/F = 10^5$ of monomer hopping rate $D$ to (per site) deposition rate $F$.   However, due to the rejection of boundary events, an extremely low parallel efficiency 
 was obtained.\cite{yshimnovotny}  Furthermore, in order to use such an approach,  in general one needs to know in advance all the possible events and their rates and then to map them to Metropolis dynamics so that all events may be selected with the appropriate probabilities.  While such a mapping may be carried out for  the simplest models, for more complicated models it is likely to be prohibitive.

A  more efficient algorithm, which is also rigorous, is the synchronous relaxation (SR) algorithm. \cite{srelax1, srelax2}  
This algorithm  was originally used by  Eick et al \cite{srelax1}  to simulate large circuit-switched communication networks and more recently by Lubachevsky and Weiss\cite{srelax2} in the context of Ising model simulations.    
In this approach, all processors remain synchronized at the beginning and end of a time interval $T$, while an iterative relaxation method is used to correct errors due to boundary events.  This algorithm has the advantage of generality (for example, it is not necessary to know the types and/or rates of all possible events in advance) and flexibility since the cycle length $T$  can be dynamically  tuned\cite{yshimrelax} to optimize the parallel efficiency.  

Recently, we have  studied the parallel efficiency and applicability of the SR algorithm in parallel KMC simulations of epitaxial growth\cite{yshimrelax}  and   have found that in some cases  a reasonable  parallel efficiency can be obtained.  
However, due to fluctuations (which increase  logarithmically\cite{yshimrelax} with the number of processors $N_p$) as well as the requirement of  global communications at the end of each cycle (the global communications time also  increases logarithmically with $N_p$) the computational speedup as a function of $N_p$ is sublinear for fixed processor size.  In addition, implementing such an algorithm is relatively complex. Therefore, there is a need for a somewhat simpler and more efficient algorithm.

In order to address these problems, we have developed a simpler synchronous sublattice (SL) algorithm for parallel kinetic Monte Carlo which we describe  in detail here.  While the SL algorithm is not  rigorous, we find that using certain reasonable assumptions on the cycle length  and  processor size, the results obtained are identical to those obtained in serial simulations.  Furthermore, because the SL algorithm requires only local  communications, the parallel efficiency is  essentially  independent  of the number of processors in the large $N_p$ limit, thus leading to linear scaling.  As a result, the parallel efficiency is  in general significantly greater than for the synchronous relaxation algorithm.

The organization of this paper is as follows.  In Section II we describe the algorithm.  In Section III we present results obtained using this algorithm for several different models of thin-film growth, including a comparison with serial results.  We also study  the effects of fluctuations on the parallel efficiency and present results for   the measured and theoretical parallel efficiency as a function of processor size and number of processors.   The effects of finite processor size on the accuracy of the results obtained using our algorithm are also discussed and compared with finite-size effects due to finite system-size.  Finally, in Section IV we summarize our results and discuss the general applicability of the SL algorithm to 
 parallel kinetic Monte Carlo simulations. 

\section {Synchronous Sublattice Algorithm}

\begin{figure}[]
\includegraphics [width=6.5cm] {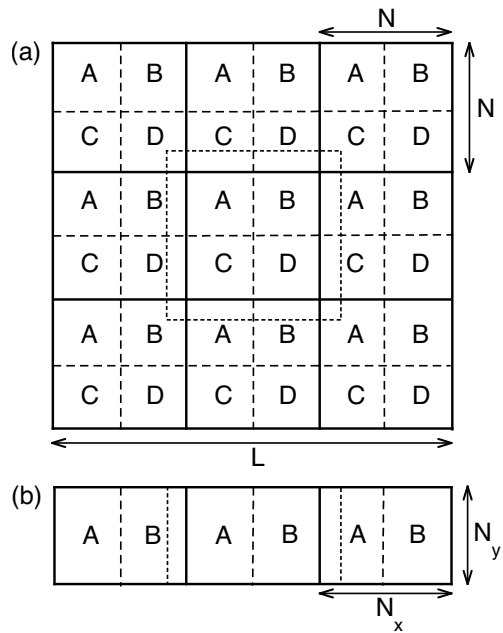}
\caption{\label{Figsub}  {Diagram showing  (a) square sublattice decomposition (9 processors) and (b) strip sublattice decomposition (3 processors).  Solid lines correspond to processor domains while dashed lines indicate sublattice decomposition. Dotted lines in (a) and (b) indicate ``ghost-region" surrounding central processor. }}
\end{figure}

\begin{figure}[]
\includegraphics  [height=4.0cm] {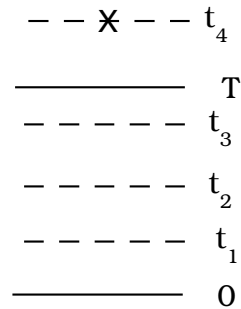}
\caption{\label{Figcycle}  { Diagram showing time-evolution in SL algorithm.  Dashed lines  correspond to selected events, while dashed line with an $\mathsf{X}$ corresponds to an event which is rejected since it exceeds the cycle time.}}
\end{figure}

As in previous work on  the ``conservative" asynchronous algorithm,\cite{Lubachevsky1, Korniss1} in  the synchronous sublattice (SL) algorithm, different parts of the system are assigned  via spatial decomposition to different processors.  However, in order to avoid conflicts between processors due to the synchronous nature of the algorithm,  each processor's domain is further divided into different regions or sublattices (see Fig. \ref{Figsub}). A complete synchronous cycle corresponding to a time interval $T$ is then  as follows.   At the beginning of a  cycle   each processor's   local  time is initialized to zero.  One of the sublattices   is then randomly selected so that all processors operate on the same sublattice during that cycle. Each processor then simultaneously and independently carries out KMC events in the selected sublattice 
until the time of the next event exceeds  the time interval $T$ (see Fig. \ref{Figcycle}).   
As in the usual serial KMC, each event is carried out with  time increment $\Delta t_i = - ln(r_i)/R_i$  where $r_i$ is a uniform random number between $0$ and $1$ and $R_i$ is the total KMC event rate.
Each processor then communicates any necessary changes (boundary events) with its neighboring processors,  updates its  event rates and moves on to the next cycle using a new randomly chosen sublattice. 

Fig. \ref{Figsub} shows two possible   methods of spatial and sublattice decomposition which are appropriate for simulations of thin-film growth$-$ a square sublattice decomposition (Fig. \ref{Figsub}(a)) and  a strip sublattice decomposition (Fig. \ref{Figsub}(b)).  In the square sublattice decomposition, the system is divided into squares, each of which is assigned to a different processor,   and each processor's domain is further divided into 4 square sublattices.  At the beginning of each   cycle one of the 4 sublattices (A,B,C, or D) is randomly chosen.  In the strip-sublattice geometry,  the system is divided into strips, each of which is assigned to a different processor,  and  each processor's domain is further divided into 2 strips or  sublattices.  At the beginning of each   cycle one of the 2 sublattices (A or B) is randomly chosen.

In order to avoid conflicts, the sublattice size must be  larger than the range of interaction (typically only a few lattice units in simulations of thin-film growth).  In addition, in order for each processor to calculate its event rates, the configuration in neighboring processors must be known as far as the range of interaction.  As a result, in addition to containing the configuration information for its own domain, each processor's array also contains a ``ghost-region" which includes the relevant information about the neighboring processor's configuration beyond the processor's  boundary.  

At the end of each cycle, each processor  exchanges information with its neighboring processors in order to properly update  its corresponding boundary  and ghost regions. For example, if sublattice A is selected in the case of square-sublattice decomposition, then at the end of a cycle, possible boundary events must be communicated to the three processors north, west and northwest of each processor.  By using sequential north and west  communications, one can eliminate the northwest  communication, and so only two communications are needed at the end of each cycle.   
Similarly, if sublattice B is selected in the case of strip-sublattice decomposition, then at the end of a cycle, possible boundary events must be communicated to the processor to the east.

Since moves are only allowed in the selected sublattice during a cycle, several cycles are needed for the entire system time to progress by $T$.   Thus, in the square (strip) geometry, it takes on average $4$  cycles ($2$ cycles) to increase the overall system time by $T$. During each cycle, the  event rates in the  non-selected sublattices of a given processor  are automatically updated as each event proceeds, just as in the usual serial KMC. 
Sublattice selection  can be carried out either by having one processor select the sublattice for that cycle and then distribute it to all processors, or more efficiently by seeding all processors with the same random number generator  so that they all independently select the same sublattice for each cycle.

We note that due to the reduced communication in the strip-sublattice decomposition compared to the square-sublattice decomposition,   the strip-sublattice decomposition is more efficient. In addition, since the sublattice in the strip-geometry is twice as large as for the square-geometry for the same processor size $N_x N_y$, there will be twice as many events per cycle in the strip geometry thus further reducing the overhead due to communication time. Thus, we expect that the  overhead due to communication latency in the strip-geometry will be approximately one-half of that for the square-geometry.

We now consider the validity and efficiency of the synchronous  sublattice (SL) algorithm.  If the time interval $T$ is not too large, then the SL algorithm  corresponds to allowing different sublattices to get slightly  ``out of synch" during each cycle.  Over many cycles one expects such fluctuations to cancel out and so the parallel evolution should be identical to the corresponding serial KMC simulation.  Of course, in order to maximize the efficiency of the algorithm (i.e. the average number of events per processor per cycle) and minimize the communication time overhead, one would like to have the largest possible value of $T$ which does not  ``corrupt" the time-evolution. 
As we shall demonstrate below, by picking the cycle length $T$ less than or equal to the average time for the fastest possible activated event (e.g. monomer hopping in the simplest possible model of thin-film growth) we do indeed obtain 
(except for very small processor sizes for which finite-size effects may occur) 
results which are identical to those obtained in serial KMC except for very small sublattice sizes.  Thus, by using the general rule that the time interval $T$ must be smaller than or equal to the inverse of   the fastest possible event rate in the KMC table, we expect that the synchronous algorithm will provide accurate results for sufficiently large sublattices.  We note that  the synchronous sublattice algorithm can also be used in a ``self-learning" KMC\cite{Trushin}  in which the KMC rate-tables are updated as the simulation goes along.  In this case, if a new ``fastest-event-rate" is discovered in the middle of a cycle, then one merely restarts the cycle from the beginning using a smaller cycle time $T$. 

\section {Results}

\begin{figure}[]
\includegraphics  [height=3.3cm] {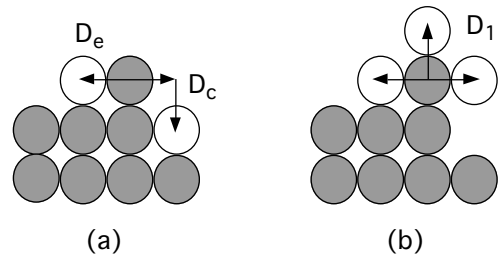}
\caption{\label{Figrelax}  {Schematic diagram of  island-relaxation mechanisms for (a) edge-and-corner  and (b)  reversible models.}}
\end{figure}

In order to test the performance and accuracy of our synchronous sublattice  algorithm we have used it to simulate three specific models of thin-film growth. In particular, we have studied  three  solid-on-solid (SOS)   growth models on a square lattice:  a ``fractal" growth model, an edge-and-corner diffusion (EC) model, and a reversible model with one-bond detachment (``reversible model").  In each of these three models the lattice configuration is represented by a two-dimensional array of heights and periodic boundary conditions are assumed.  In the ``fractal" model,\cite{islandprb}  atoms (monomers) are deposited onto a square lattice with (per site)  deposition rate $F$, diffuse (hop) to nearest-neighbor sites with hopping rate $D$ and attach irreversibly  to other monomers or clusters via a nearest-neighbor bond (critical island size of $1$).  The key parameter  is the ratio $D/F$  which is typically much larger than one in epitaxial growth.   In this model fractal islands are formed in the submonolayer regime  due to the absence of island relaxation. The EC model is the same as the fractal model except that island relaxation is allowed, i.e.   atoms which have formed a single nearest-neighbor bond with  an  island may diffuse along the  edge of the island with diffusion rate $D_e = r_e D$ and around island-corners with rate $D_c = r_c D$ (see Fig. \ref{Figrelax}).  Finally, the reversible model is  also similar to the fractal model except that  atoms with one-bond (edge-atoms) may hop along the step-edge or away from the step with rate $D_1 = r_1 D$, thus allowing both edge-diffusion and single-bond detachment.  For  atoms hopping up or down a step, an extra Ehrlich-Schwoebel barrier   to interlayer diffusion \cite{ES}  may also be included.  In this model, the critical island size  $i$\cite{Venables} can vary from $i = 1$ for small values of $r_1$,  to $i = 3$ for  sufficiently large values of  $D/F$ and  $r_1$.\cite{islandprl}

 For the fractal and reversible models, the range of interaction corresponds to one nearest-neighbor (lattice) spacing, while for the EC model it corresponds to the next-nearest-neighbor distance.  Thus, for these models  the width of the ``ghost-region"  corresponds to one   lattice-spacing.  We note that at each step of the simulation, either a particle is   deposited within the appropriate sublattice, or a particle  diffuses  to a nearest-neighbor or next-nearest-neighbor lattice site.  In order to avoid ``double-counting",  only particles within a processor's domain may diffuse, e.g.  if a particle  diffuses from the boundary region of a processor into its ghost-region during a cycle, then that particle is no longer free to move during that cycle.  In more general models, for which concerted moves involving several atoms may occur, \cite{Zhangdimer, Hamilton, Jonsson, tad4} the ghost region needs to be at least as large as the range of interaction and/or the largest possible concerted move.  In such a case, the processor and sublattice to which a concerted event belongs can be determined by considering the location of the center-of-mass of the atoms involved in the concerted move. 
 
 In order to maximize both the serial and parallel efficiency  in our KMC simulations,  we have used  lists to keep track of all possible events of each type and rate.  For each sublattice, a  set of lists is maintained which contains  all possible moves of each type.  A binary tree  is used to select which type of move will be carried out, while the particular move is then randomly chosen from the list of the selected type.  After each move, the lists are updated. 

\subsection {Computational Details}

In order to test our algorithm we have carried out both ``serial emulations" as well as    parallel simulations. However, since our main goal is to test the  performance and scaling behavior on parallel machines we have primarily focused on direct parallel simulations using the Itanium and AMD clusters at the Ohio Supercomputer Center (OSC) as well as on the Alpha cluster at the Pittsburgh Supercomputer Center (PSC).  All of these clusters have fast communications$-$the Itanium and AMD clusters have Myrinet and the Alphaserver cluster has Quadrics. In our simulations, the interprocessor communications  were carried out using MPI (Message-Passing Interface).  

 \subsection {Comparison with Serial Results}
 
\begin{figure}[]
\includegraphics [width=6.0cm] {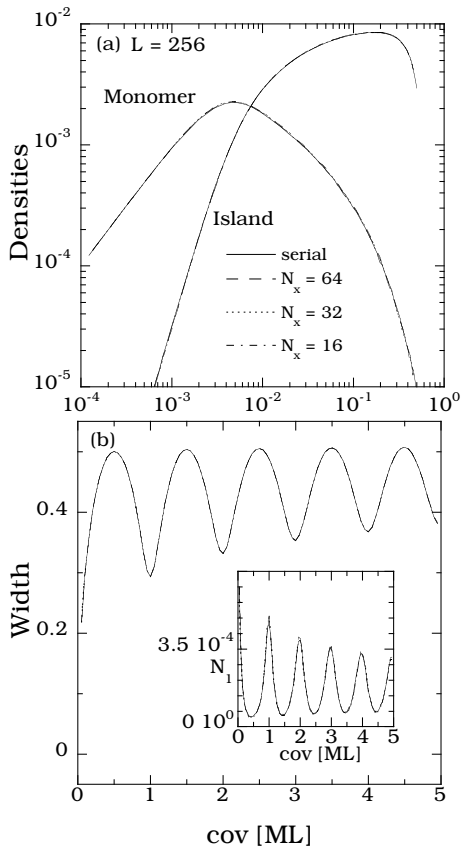}
\caption{\label{Figaccuracy1}  {Comparison between serial and parallel results using synchronous sublattice algorithm with strip decomposition ($L = N_y$) for  fractal model with $D/F = 10^5$.}}
\end{figure}

\begin{figure}[]
\includegraphics [width=6.0cm] {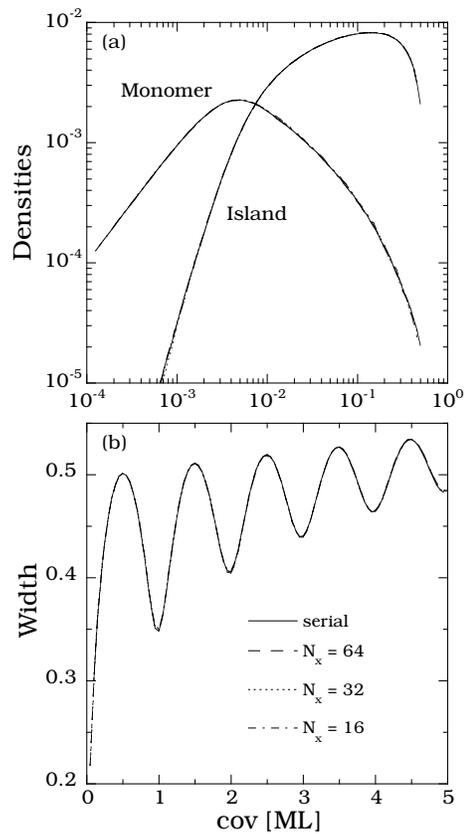}
\caption{\label{Figaccuracy2}  {Comparison between serial and parallel results using synchronous sublattice algorithm with strip decomposition for   EC model with $D/F = 10^5$, $L = 256$, and $D_e = 0.1 D$, $D_c = 0$.}}
\end{figure}

As a test of our algorithm we first present some detailed comparisons with serial results for different numbers of processors and system sizes for both the square and the strip geometries. 
Fig. \ref{Figaccuracy1} shows  a  comparison of parallel and  serial results for the fractal model with $D/F = 10^5$ and a square system of  size $L = 256$.  The parallel simulations were carried out using   a strip sublattice decomposition with  processor sizes $N_x = 16, 32$ and $64$  with $N_y = 256$ corresponding to $N_p =  16, 8,$ and $4$  respectively, where $N_p$ is the number of processors.  In particular, Fig. \ref{Figaccuracy1}(a) shows   the substrate monomer density $N_1$ and island density $N$  (averaged over 500 runs) as a function of coverage  in the first half-layer of growth, while Fig. \ref{Figaccuracy1}(b) shows the r.m.s. surface height fluctuations or surface width  (averaged over 100 runs)  as a function of coverage in the first few layers of growth.  The inset of Fig. \ref{Figaccuracy1}(b) also shows the monomer density as a function of coverage in the first 5 layers of growth.  As can be seen, there is no difference 
within statistical error between the serial and the parallel results.  
A similar comparison  is shown in Fig. \ref{Figaccuracy2} for the  edge-diffusion (EC)  model ($D/F = 10^5, r_e = 0.1, r_c = 0$) using  a strip sublattice decomposition.  As can be seen there is again no difference between the parallel  and serial results. 

\subsection {Parallel Efficiency as a Function of $D/F$}

We now consider the performance of the synchronous sublattice algorithm, starting with the dependence of the parallel efficiency  on the monomer diffusion rate $D/F$ for the fractal model for a fixed number of processors ($N_p = 4$).  
Here we  define the parallel efficiency  as equal to the ratio of the execution time for an ordinary serial simulation of one processor's domain (using the  same sublattice decomposition as in the parallel simulation) to  the parallel execution time of $N_p$ domains using $N_p$ processors.  Thus, the overall ``performance factor" of the parallel simulation (e.g. boost in events/sec over a serial simulation) is given by the parallel efficiency multiplied by the number of processors $N_p$. 

There  are two primary   factors which determine  the parallel efficiency.  The first is the overhead due to communications at the end of every cycle, when all processors exchange boundary information with their neighbors.  Since in our simulations the number of boundary events is relatively small (i.e. the processor size is not too large) the primary cause of   communications overhead is the latency time for local communications  which is independent of processor domain size.   The second important factor controlling the efficiency is the existence of fluctuations in the number of events in different processors.  In particular, in any given cycle one processor may have many events, while its nearest-neighbor  may have fewer events.  As a result, while the processor with many events is calculating its events, its neighboring processor with few events must idle (wait) until it has received the boundary information from the first processor before moving to the next cycle.

To illustrate this effect more quantitatively, we consider the effects of fluctuations on the parallel efficiency in the case of the one-dimensional strip sublattice decomposition shown in Fig. \ref{Figsub} (b). In this case there are two sublattices (A and B) and during each cycle one of the sublattices is randomly selected.  For example, if the B sublattice is selected, then at the end of a cycle all processors will do a (non-blocking) send of any boundary events in the B sublattice  to the processor on their right, followed by a (blocking) receive of the corresponding boundary information from the processor on their left.  Thus, for example, if processor 1 has more events than processor 2, and so takes longer to execute these events before initiating its send to processor 2, then processor 2 must wait before moving to  the next cycle, thus leading to inefficiency. However, processor 2's execution  is not affected by processor 3 during the same cycle, since its send to processor 3 is non-blocking.  

Denoting the communication overhead   per  cycle  as $t_{com}$  and taking into account the fluctuations of events between nearest-neighbor processors, we obtain the following  expression for the average time per cycle:
\begin{equation}
                     t_{av} (N_p)  =     t_{1p} + t_{com} +    \langle \Delta(\tau) \rangle ~ (t_{1p}/n_{av})  \label{cycletime}
 \end{equation}
where  $t_{1p}$ is the time for a single processor serial simulation of a single processor's domain, $n_{av}$ is the average number of events per processor per cycle, $ \Delta (\tau)$ is the relevant  fluctuation in a given cycle $\tau$ (averaged over all processors), and the brackets denote an average over all cycles. The ratio $t_{1p}/n_{av}$ in the last term of Eq.~1 corresponds to the average calculation time to process an event.  Therefore, the parallel efficiency $PE$ may be written as
\begin{equation}
         PE  =   \frac {t_{1p}} {t_{av} (N_p)} =  [~1 + \frac {t_{com}} {t_{1p}} +  \frac {\langle  \Delta (\tau) \rangle } {n_{av}} ~]^{-1} \label{PEsublattice}
 \end{equation}
In the limit of negligible communication time $t_{com}/t_{1p}  \rightarrow 0$, this implies that the maximum possible parallel efficiency is  given by,
\begin{equation}
        PE^{max}  =     [~1 + \frac {\langle  \Delta (\tau) \rangle } {n_{av}}~]^{-1}  \label{PEsublatticemax}
 \end{equation}

We also note that $n_{av} \sim N_x N_y $  and since the fluctuations are {\it on average} uncorrelated, one expects 
$\langle \Delta(\tau) \rangle  \sim \sqrt {n_{av}}$.  
This implies that the maximum possible parallel efficiency may be written as,
\begin{equation}
         PE^{max}  = [~1 + \frac {\alpha} {(N_x N_y )^{1/2}} ~]^{-1} \label{PEsublatticemax2}
 \end{equation}
where the constant $\alpha$ is model-dependent. This result shows  clearly that the maximum theoretical efficiency 
approaches $1$ in the limit of large $n_{av}$ corresponding to large $N_x, N_y$.

There are two distinct ways in which the average fluctuation $\Delta (\tau)$ might be calculated. If we assume that at the beginning of each cycle  all processors are perfectly synchronized, then for the strip geometry one may write,
\begin{equation}  
\Delta_S  (\tau) = \frac {1} {N_p}  \sum_{i=1}^{N_p}  ( n_{i+\delta(\tau)} (\tau) - n_{i} (\tau)) ~ \Theta (n_{i+\delta(\tau)} (\tau) - n_{i}(\tau) )\label{synch1}
\end{equation}
where $n_i(\tau)$ is the number of events in processor $i$ in cycle $\tau$, $\Theta(x) = 0 (1)$ if $x$ is negative (positive)  and $\delta(\tau)  = +1 (-1)$ if the A (B) sublattice is selected in cycle $\tau$. Since we are interested in the average over many cycles, this is equivalent to the simpler form,
\begin{equation}  
\Delta_S  (\tau) = \frac {1} {2 N_p}  \sum_{i} | n_i(\tau) - n_{i+1}(\tau) |   \label{synch}
\end{equation}
where the factor of $1/2$  is due to the fact that  only half the time will the relative fluctuation in the relevant neighboring processor be positive, and thus lead to a delay.  

However, 
due to fluctuations 
one must also take into account the existence of  desynchronization at the beginning of a cycle.  In order to take this into account, we may calculate the sum or ``starting time"  $S_i(\tau)$ corresponding to the sum of the  total number of events in processor $i$ and the sum of all delay-events due to neighboring processors in a given processor $i$ at the start of cycle $\tau$. At the start of the first cycle ($\tau = 1$) one has $S_i(1) = 0$ for all processors $i$ and $n_i(1)$ is the number of events in processor $i$ in that cycle.  At the start of each subsequent cycle, the sum $S_i(\tau)$ may be calculated in each processor in terms of the previous values of $S_i(\tau-1)$ and $n_i(\tau - 1)$  as follows,
\begin{equation}
S_i(\tau) = S_i(\tau-1) + n_i(\tau -1) + \Delta_i(\tau) \Theta(\Delta_i(\tau)) \label{recursion}
\end{equation} 
where 
\begin{equation}
\Delta_i(\tau) = S_{i+\delta(\tau)}(\tau - 1) +  n_{i+\delta(\tau)}(\tau-1)  - S_{i}(\tau - 1) - n_{i} (\tau -1)
\end{equation}
and where $\delta(\tau)  = +1 (-1)$ if the A (B) sublattice is selected in cycle $\tau$. 
Then the average delay $\Delta(\tau)$ due to fluctuations in a given cycle $\tau$ may be written,
\begin{equation}
\Delta (\tau) = \frac {1}{N_p} \sum_{i=1}^{N_p} \Delta_i(\tau) \Theta (\Delta_i(\tau))
\end{equation}

\begin{figure}[]
\includegraphics [width=8.5cm] {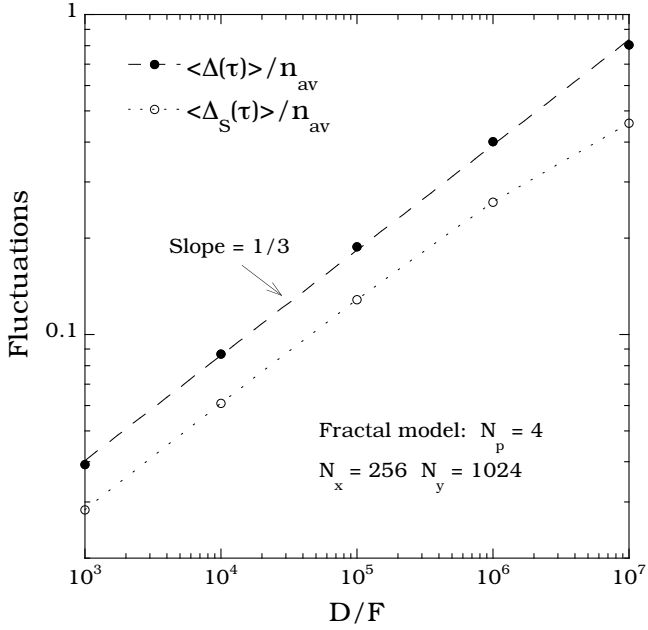}
\caption{\label{Figflucdf}  {Fluctuations as function of $D/F$ for fractal model with $N_p = 4$ and strip geometry with $N_x = 256$, $N_y = 1024$.}}
\end{figure}

Fig. \ref{Figflucdf} shows the measured fluctuations  $\langle \Delta(\tau)\rangle/n_{av}$ and $\langle \Delta_S(\tau)\rangle/n_{av}$  for the simple fractal model as a function of  $D/F$ for fixed processor size $N_x = 256$, $N_y = 1024$ and $N_p = 4$.  As can be seen, for $N_p = 4$, the full fluctuation $\langle \Delta (\tau)\rangle$ is   approximately 30\% larger than that obtained assuming that all processors are synchronized at the beginning of each cycle.      
For the simple fractal model,   one expects  $n_{av} \sim N_1  \sim (D/F)^{-2/3}$ which implies $ \langle \Delta(\tau) \rangle/n_{av}   \sim (D/F)^{1/3}$. As can be seen in Fig. \ref{Figflucdf},   there is very good agreement with this form for the $D/F$-dependence. 

\begin{figure}[]
\includegraphics [width=8.5cm] {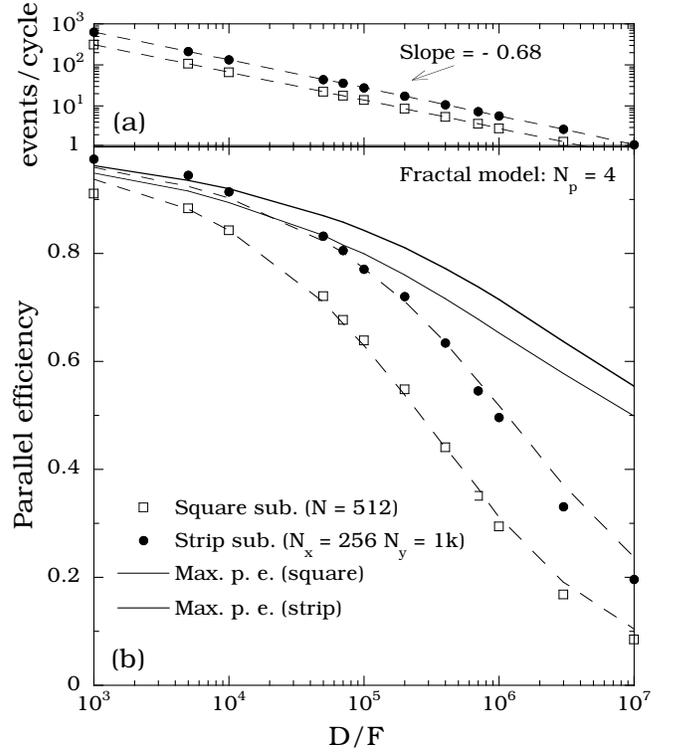}
\caption{\label{Figfracpedf}  {(a) Events per cycle and (b) parallel efficiency for fractal model with $N_p = 4$  as function of D/F.  Dashed lines correspond to Eq. \ref{PEsublattice} while solid lines correspond to maximum theoretical efficiency given by Eq. \ref{PEsublatticemax}. }}
\end{figure}

Fig. \ref{Figfracpedf} shows the corresponding results (symbols) for the parallel efficiency as a function of the ratio $D/F$.  Results are shown for parallel KMC simulations with $N_p = 4$ of a square system with system size $L = 1024$ with both square sublattice decomposition ($N_x = N_y = 512$) and strip sublattice decomposition ($N_x = 256, N_y = 1024$). Due to the decreased communication overhead in the strip-geometry (1 send/receive versus 2 send/receives) the parallel efficiency of the strip geometry simulations is significantly larger than for the square geometry. 
As can be seen, for $D/F \le 10^6$, the parallel efficiency for the strip geometry is greater than 50\%.  However, with increasing $D/F$ the parallel efficiency decreases significantly since the decrease in the  number of events per cycle $n_{av}$  (see Fig. \ref{Figfracpedf} (a)) leads to an increase in the communications overhead  $ {t_{com}}/ {t_{1p}}$ as well as  in the relative fluctuations   ${\langle  \Delta (\tau) \rangle }/ {n_{av}}$.

Also shown  in Fig.~\ref{Figfracpedf}   (dashed lines) is  the  parallel efficiency  calculated using Eq.~2 based on   the measured values of  $\langle \Delta (\tau) \rangle/n_{av}$, $t_{1p}/n_{av}$, and the measured interprocessor communication time $t_{com} \simeq 15\mu s$ per send/receive. As can be seen, there is good agreement between the measured and calculated results. The maximum theoretical parallel efficiencies calculated using Eq.~3  assuming  negligible  communication overhead  are also shown (solid lines).  As can be seen, the maximum theoretical efficiencies are significantly higher than the measured efficiencies for large $D/F$, although they also decrease with increasing $D/F$ due to the increase in fluctuations. For the simple fractal model with strip-geometry, our results for the maximum possible parallel efficiency  may be well described by the expression, 
\begin{equation}
        PE^{max}_{frac}   = [~1 + 0.21 (D/F)^{1/3}/(N_x N_y )^{1/2} ~]^{-1} \label{PEsublattice3}
 \end{equation}
This result may be used to estimate the maximum possible efficiency for the fractal model for different processor sizes and values of $D/F$.  For example, if $N_x = N_y = 64$ and $D/F = 10^5$, then this implies a maximum possible parallel efficiency given by $PE^{max} = 0.40$. This result shows  that even in the absence of delays due to interprocessor communication, due to the existence of fluctuations, the   parallel efficiency will decrease with increasing   $D/F$.

\begin{figure}[]
\includegraphics [width=8.5cm] {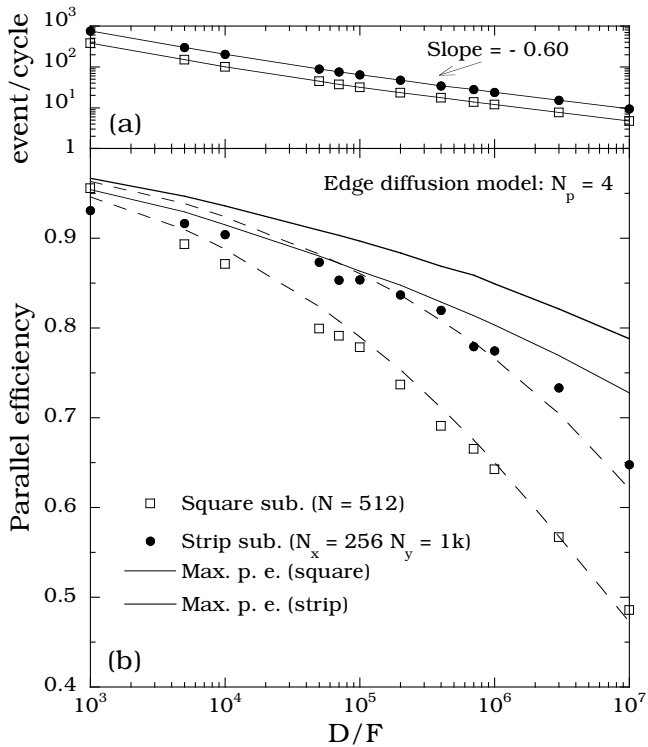}
\caption{\label{Figedgepedf}  {(a) Events per cycle and (b) parallel efficiency for edge-diffusion model with $N_p = 4$ as function of $D/F$.  Dashed lines correspond to Eq. \ref{PEsublattice} while solid lines correspond to maximum theoretical efficiency given by Eq. \ref{PEsublatticemax}. }}
\end{figure}

Fig. \ref{Figedgepedf} shows similar results for the parallel efficiency as a function of   $D/F$ for  the edge-diffusion model with $D_e = 0.1 D, D_c = 0$, $N_p = 4$ and $\theta=1ML$.  As can be seen, although the parallel efficiency for the edge-diffusion model still decreases with increasing $D/F$,  it is significantly larger than for the fractal model.   In particular,  due to the increased number of events per cycle and  the resulting reduced communication overhead, the parallel efficiency remains above 50\% for large $D/F$.  As an example, for the case of strip-geometry and $D/F = 10^7$, the parallel efficiency is more than  3 times that for the simple fractal model, while the number of events is approximately 10 times larger.  As for the fractal model, the calculated parallel efficiency (dashed lines)  is in good agreement with the  measured values. Due to the increase in the number of events per cycle, and the resulting decrease in the relative fluctuations, the maximum theoretical efficiencies (solid lines) are also significantly higher than for the fractal case.

\subsection {Parallel efficiency as function of  number of processors for fixed processor size}

\begin{figure}[]
\includegraphics [width=8.5cm] {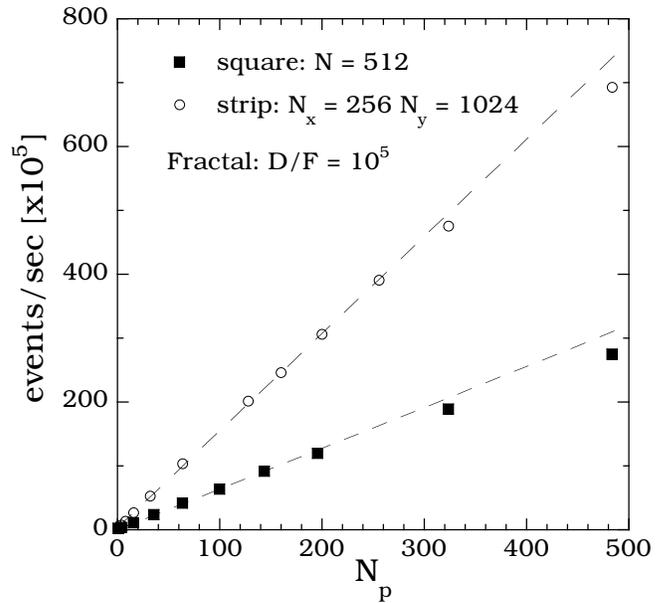}
\caption{\label{Figcompspeed}  {Total computational speed (events/sec) as function of number of processors for fractal model with $D/F = 10^5$.}}
\end{figure}

Fig. \ref{Figcompspeed} shows  the performance (events/sec)  for the simple fractal model with $D/F = 10^5$ as a function of the number of processors $N_p$ with fixed processor size,  using both  square decomposition with $N = 512$ and strip decomposition with $N_x = 256$ and $N_y = 1024$.  
As can be seen in both cases there is a roughly linear speedup of the performance with increasing number of processors $N_p$.   For comparison,  the equivalent single-processor (serial) computation speed for this model is  approximately $2.8 \times 10^5$ events/sec.  However, due to the decreased communication cost, the speed-up using the strip geometry is significantly higher than for the square decomposition.

\begin{figure}[]
\includegraphics [width=8.5cm] {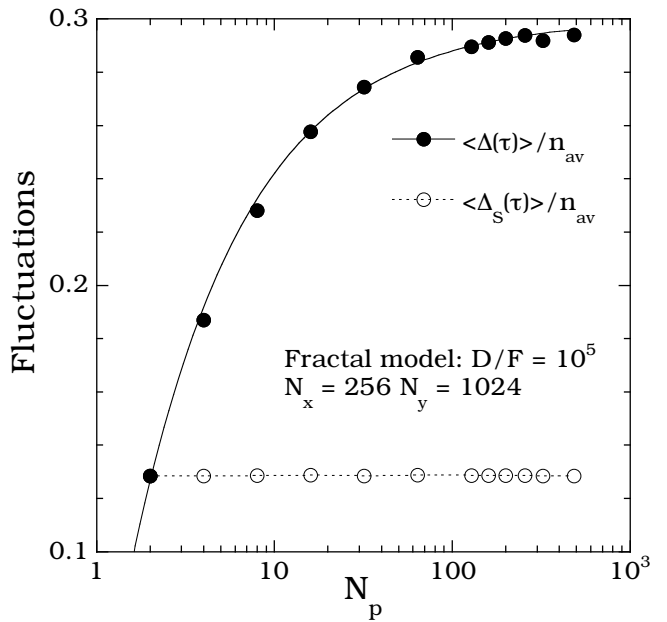}
\caption{\label{Figflucnp}  {Fluctuations for fractal model with $D/F = 10^5$ as function of $N_p$.  Solid line corresponds to fit of form $\langle \Delta_S(\tau)\rangle/n_{av} = 0.3 - 0.28/N_p^{0.68}$}}
\end{figure}

We now consider the dependence of the parallel efficiency on the number of processors $N_p$ in more detail for the case of strip geometry.  Fig. \ref{Figflucnp} shows the measured fluctuations  $\langle \Delta(\tau)\rangle/n_{av}$ and $\langle \Delta_S(\tau)\rangle/n_{av}$ as a function of $N_p$ for the fractal model for $D/F = 10^5$.  With increasing $N_p$, the  relative event-fluctuation  $\langle \Delta_S(\tau)\rangle/n_{av}$ remains constant.  In contrast,  the  full fluctuation $\langle \Delta(\tau)\rangle/n_{av} $ increases slowly with increasing $N_p$ but appears to saturate at a finite value 
for large $N_p$. This is supported  by the  fit shown in Fig. \ref{Figflucnp} (solid line)  which    agrees quite well with the simulation results and which has the form, $\langle \Delta_S(\tau)\rangle/n_{av} = 0.30 - 0.28/N_p^{0.68}$.  
Due to the saturation of fluctuations, we expect that the parallel efficiency will also saturate for large $N_p$.

\begin{figure}[]
\includegraphics [width=8.5cm] {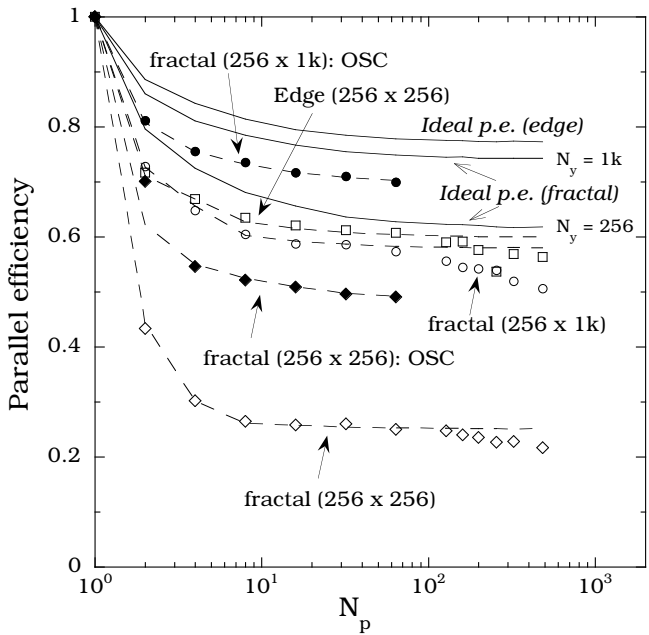}
\caption{\label{Figpenp}  {Parallel efficiency (symbols) as function of number of processors $N_p$ for fractal and edge-diffusion models with $D/F = 10^5$ and strip-sublattice decomposition ($N_x = 256$ with $N_y = 256$ and $1024$). Dashed lines correspond to Eq. \ref{PEsublattice} while solid lines correspond to maximum theoretical efficiency calculated using Eq. \ref{PEsublatticemax}.  }}
\end{figure}

Fig. \ref{Figpenp}  shows our results for  the   measured parallel efficiency (open and closed symbols) as a function of $N_p$ for fixed processor size for the fractal and edge-diffusion models.  
As expected,  the parallel efficiency is essentially constant for large $N_p$, although there is a slight  decrease  due to  increased communication overhead  for $N_p > 100$.  Also shown (dashed lines) are  the   parallel efficiencies calculated from  Eq. \ref{PEsublattice}  using  the measured fluctuations and communication times, as well as  the maximum possible theoretical  parallel efficiencies (solid lines)  calculated using Eq. \ref{PEsublatticemax}.   As can be seen, there is relatively good agreement between the calculated and measured parallel efficiencies.

We note that the results for large $N_p$ (open symbols) were  obtained using the Alpha cluster at the Pittsburgh Supercomputer Center (PSC) for which the communication latency   is somewhat larger than for the OSC cluster.  As a result the parallel efficiencies  are somewhat lower than would be obtained with the OSC cluster.  
For comparison, OSC results  for the fractal model   with $N_x = N_y = 256$ and  with $N_x = 256, N_y = 1024$ are also shown up to $N_p = 64$ (filled symbols). As can be seen, due to the decreased communication time, the OSC results for the parallel efficiency are significantly larger than the corresponding PSC results.   
Except for the PSC fractal results with $N_x = N_y = 256$, the parallel efficiencies are all  larger than 50\%.

\subsection {Finite-Size Effects}

\begin{figure}[]
\includegraphics [width=8.5cm] {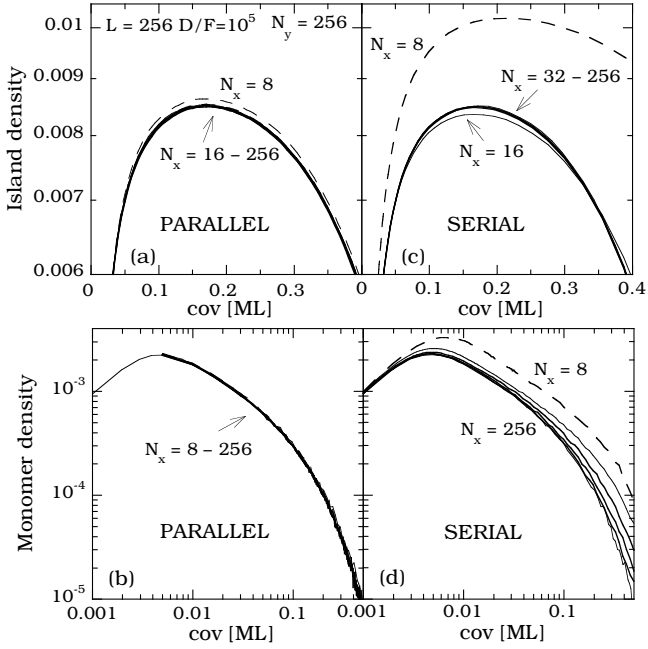}
\caption{\label{Figfse1}  {Finite-size effects in parallel and serial simulations  of fractal model with $D/F = 10^5$.   Parallel simulations (averaged over 200 runs) are for system size  $L =  256$ with processor sizes $N_x = 8, 16, 32, 64$, and $128$ and $N_y = 256$.   Serial simulations  (averaged over 500 runs) are for  system size $N_x \times 256$. }}
\end{figure}

We now consider the effects of finite processor size on the accuracy of the results obtained using the synchronous sublattice algorithm.  For simplicity we focus on the case of  strip-sublattice decomposition.   
As we have already shown (see Figs. \ref{Figaccuracy1} and \ref{Figaccuracy2}), for  sublattice sizes  which are not too small, there is perfect agreement between the synchronous sublattice results and the corresponding serial results. However,   for very small processor sizes there exists  a small    ``finite-size" effect which leads to results which are slightly different from those obtained using the usual serial KMC algorithm. In particular, as shown in Fig. \ref{Figfse1},   there is essentially perfect agreement between the synchronous sublattice results  for the fractal model with  system size $L = 256$,  $D/F = 10^5$, and $N_x = 16 - 256$  and the corresponding serial results.  However, for the smallest processor size ($N_x =   8$)  there is approximately a 2\% difference between the synchronous sublattice  results for the peak island density $N$ and the corresponding  serial results (although there are no differences in the monomer density $N_1$).   In order to compare  these effects  with those of finite system size, in Fig. \ref{Figfse1} (c) and (d) we also show  the corresponding serial results for systems of size $N_x = 8 - 256$ and $N_y = 256$.  As can be seen, the finite-size effects which occur for small system size are much  larger  than those  due to  finite processor size. This indicates that  the effects of finite processor size are both qualitatively different as well as much weaker than  those due to finite-system size. 

\begin{figure}[]
\includegraphics [width=8.5cm] {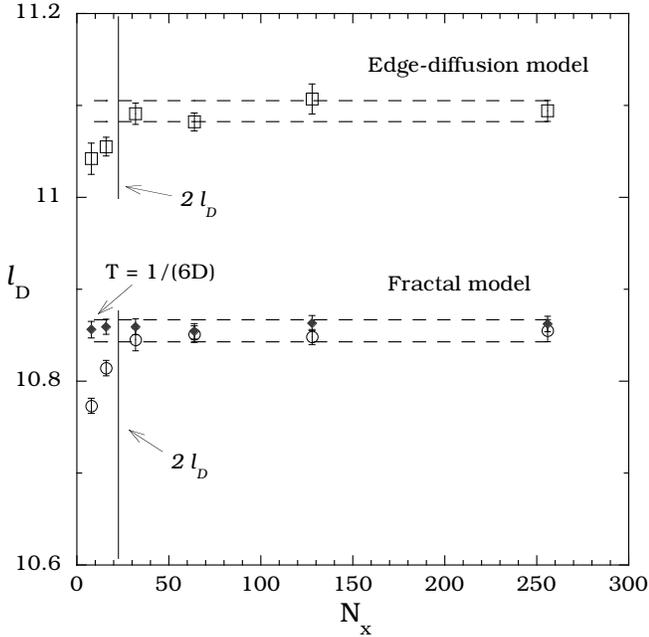}
\caption{\label{Figfse2}  {Diffusion length  $l_D = N_{pk}^{-1/2}$ in parallel and serial simulations  of fractal model (200 runs) and edge-diffusion model  (100 runs) with $D/F = 10^5$ and $T = 1/D$ (open symbols).   All simulations
are for system size  $L = 256$ with processor sizes $N_x = 8, 16, 32, 64, 128$, and $256$ and $N_y = 256$.   Horizontal dashed lines correspond to error bars for serial simulations. }}
\end{figure}

We now consider  these finite-size effects in somewhat more detail. While a variety of   length scales (such as the typical mound or feature size in multilayer growth)  may occur in the models studied here, there is  one {\it dynamical}  length scale corresponding to the ``diffusion length" $l_D$ (e.g. the typical distance a monomer may diffuse before being captured)  which plays a particularly important role.    In particular, the diffusion length may be written in terms of the peak submonolayer island density, i.e. $l_D \sim  N_{pk} ^{-1/2}$.  Since in the synchronous sublattice algorithm, particles which diffuse outside the active sublattice  are no longer active during that cycle,  we conjecture that for sublattice  sizes $(N_x/2)$ which are smaller than  the  diffusion length $l_D$,  finite-size effects   may occur.  
For the fractal model studied here, one has $N_{pk} \sim (D/F)^{-1/3}$ which implies $l_D \sim (D/F)^{1/6}$, e.g. the diffusion length increases slowly with increasing $D/F$.  As shown in Fig. \ref{Figfse2}, by measuring the peak island density for  $D/F = 10^5$, we  obtain  $l_D \simeq 11$ which implies a critical processor size $N_x$ given by $N_x  \simeq 2~ l_D  \simeq 22$. This result is in good agreement with the observation of the onset of significant   finite-size effects for $N_x < 16$. 

Also shown in Fig. \ref{Figfse2} are similar results for the edge-diffusion model with $D/F = 10^5$ and $r_e = 0.1$.  In this case the diffusion length is slightly higher than for the fractal model.  However, again   there are no finite-size effects for $N_x > 2  l_D$.  Similar results have also been obtained for the  reversible one-bond detachment model, as well as a reversible bond-counting model (not shown).  In all cases, we find that there are no differences between the serial results and the parallel KMC results as long as $N_x > 2 l_D$.  In contrast, for $N_x < 2 l_D$, noticeable but weak finite-size effects are  observed. 

While these results are for a cycle length  $T = 1/D$ given by the inverse of the fastest  hopping rate,  for a smaller cycle-length we expect that    the critical processor size $N_x$ corresponding to finite-size effects will be significantly reduced.  
As shown in  Fig.  \ref{Figfse2} (filled symbols)  for the fractal model with cycle length $T = 1/(6D)$,  the critical processor size $N_x$ is significantly smaller than the diffusion length $l_D$.   However, for such a reduced cycle length, the parallel efficiency is also   significantly reduced. 

\begin{figure}[]
\includegraphics [width=8.5cm] {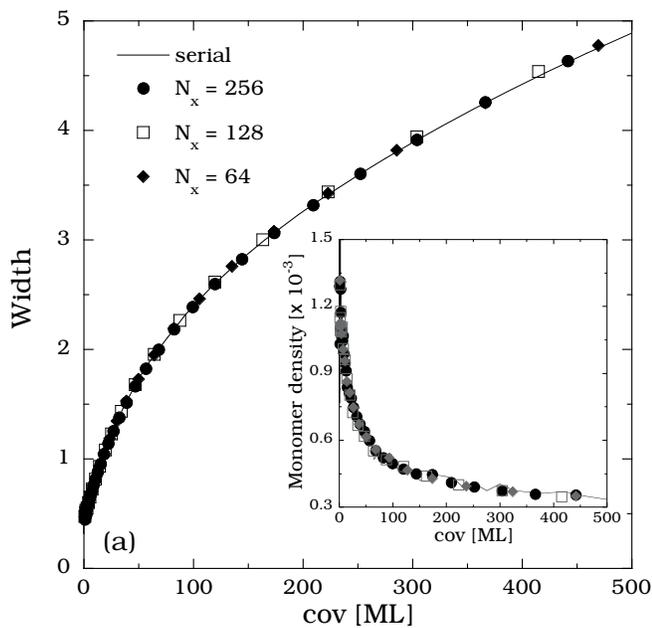}
\vskip 0.25 truein
\includegraphics [width=6.5cm] {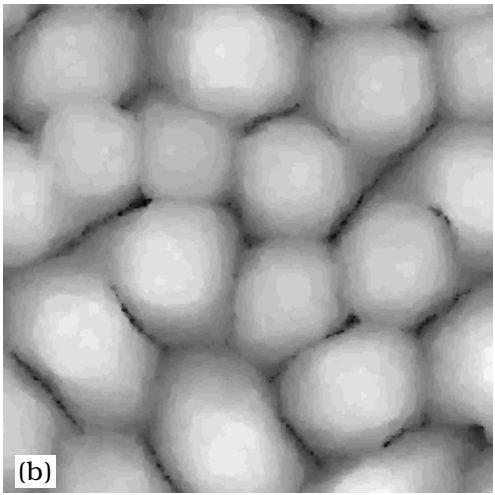}
\caption{\label{Figmultilayer}  {(a) Comparison between serial and parallel results for  reversible model with $L = N_y = 1024, D/F=10^5, E_1=0.1$ eV and $E_b=0.07eV$ ($N_p=L/N_x$). (b) Gray-scale plot of  $512 \times 512$ portion of system at $\theta = 500$ ML. }}
\end{figure}

As a further test of both  the parallel efficiency and  finite-size effects in  the SL algorithm, we have 
carried out multilayer simulations of the  reversible  model at $T = 300$ K, with system size $L = 1024$,  $D/F = 10^5$, $E_1 = 0.1$ eV, and an Ehrlich-Schwoebel barrier to interlayer diffusion given by $E_b = 0.07$ eV.  In our simulations particles freshly deposited near step-edges are assumed to ``cascade" randomly to the nearest-neighbor sites below (``knockout"). Fig \ref{Figmultilayer} (a) shows  serial results (solid line) for the r.m.s. surface height fluctuations (surface width) and monomer density  up to 500 ML along with   the corresponding parallel results obtained using the SL algorithm with processor sizes $N_y = 1024$ and $N_x = 64, 128$, and $256$ corresponding to $N_p = 16, 8$, and $4$ respectively.  As can be seen, there  is no difference between the serial and parallel results even though  the typical mound size of approximately $100$ lattice units   (see Fig. \ref{Figmultilayer}(b))  is significantly larger  than the smallest sublattice size $N_x/2 = 32$.  This indicates that the relevant length-scale determining the existence of  finite-size effects in the SL algorithm is indeed the diffusion length $l_D$ and not   the characteristic feature size.  Since in these simulations the total system size $L$ was fixed,  the parallel efficiency may be written,
\begin{equation}
         PE  =   \frac {t_{1p}} {N_p ~t_{av} (N_p)}  \label{PEsublattice2}
 \end{equation}
where $t_{1p}$ is the calculation time for a serial simulation of the entire system.   
Since the processor size decreases with increasing $N_p$,  both the relative magnitude  of  event fluctuations and  the   overhead due to communication latency will also increase.  As a result,    the parallel efficiency  decreases   with increasing $N_p$ rather than saturating as in the case of fixed processor-size.  The parallel efficiencies obtained in these simulations  were 92\% ($N_p = 4$),  85\% ($N_p = 8$), and 70\% ($N_p = 16$) respectively. 

\section {Discussion}

We have  developed and tested  a  synchronous sublattice  (SL) algorithm for parallel kinetic Monte Carlo simulations. 
In our algorithm, the maximum cycle length $T$ is given by the inverse of the fastest diffusion rate.  
For sublattice sizes which are smaller than the diffusion length $l_D$,  weak finite-size effects are observed which lead to deviations from the results obtained using a serial algorithm. However, for sublattice  sizes larger than the diffusion length $l_D$, the results obtained are identical to those obtained in serial simulations.  Since in many systems of interest the diffusion length is typically relatively small (e.g. of the order of a few to a few tens of lattice spacings) while significantly larger system sizes are  needed  to  avoid finite system-size effects,  the SL algorithm should provide a  useful, efficient, and accurate   method to carry out parallel KMC simulations of these systems. 

We have also measured the parallel efficiency of the SL algorithm as a function of the number of processors $N_p$ for fixed processor size.  Because the SL algorithm is   synchronous, the parallel efficiency is affected by   fluctuations in the number of events in different  processors over a given cycle.  However, because  only local  communications are required,  these fluctuations saturate  as the number of processors increases.  As a result, linear scaling behavior for the total speedup as a function of the number of processors is observed, e.g.  the parallel efficiency is independent  of the number of processors in the large $N_p$ limit. 

For the simple models we have studied here,  the calculation time for a single  event such as diffusion or deposition is significantly smaller than  the   latency time for nearest-neighbor communication.  As a result,  the parallel efficiency increases  with  processor size, since the communications overhead per event is reduced by the increased number of events in a cycle. However, even for relatively modest processor sizes, we have obtained reasonable values for  the asymptotic parallel efficiency $PE$  ranging from  50\% for the fractal model with $D/F = 10^5$ and $N_x = N_y = 256$, to 70\%  for the fractal model with $N_x = 256$, $N_y = 1024$.   For the  slightly more complicated edge-diffusion (EC) model, for which   the number of events per cycle is  larger,   significantly larger efficiencies are obtained for the same processor size, e.g. 60\% for $N_x = N_y = 256$ and $D/F = 10^5$.  Similarly, for the reversible model, we have obtained a parallel efficiency $PE \simeq 70\%$ for the same effective processor size ($N_x  = 64, N_y = 1024$) with $N_p = 16$.

We have also studied the effects of fluctuations  on the parallel efficiency in our simulations. In particular, we  found that the relevant relative fluctuations $\langle \Delta_S(\tau)\rangle/n_{av}$ scale as  one over the square root of the processor size (see Eq.\ref{PEsublattice3} and Fig.\ref{Figflucdf}). By taking  into account  the effects of fluctuations and communication delays, calculated parallel efficiencies were obtained which are in excellent agreement with those obtained in our simulations.   In addition, by measuring the relevant fluctuations, we have been able to predict the maximum possible theoretical efficiencies in the absence of communication delays. For example,  for the fractal and edge-diffusion models with $D/F = 10^5$ and $N_x = N_y = 256$, maximum theoretical parallel efficiencies of 80\% and 90\% respectively were obtained.  

 Since  increasing the processor size decreases the effects of fluctuations as well as  communications overhead,   the parallel efficiency may be further  increased by increasing the   processor size. Alternatively, in simulations on machines with faster communications (such as shared memory machines)    or   in simulations of more complicated KMC models   for which the calculation time is significantly larger (such as self-learning KMC\cite{Trushin} or accelerated dynamics\cite{tad4})    efficiencies approaching these values may be possible even without increasing the processor size.

It is worth noting that in our simulations we have used two slightly different definitions for  the  parallel efficiency. 
In the first definition (Eq. \ref{PEsublattice}),  the parallel execution time was compared with  the serial execution time of a system whose size is the same as a single processor.  In contrast, in the second definition (Eq. \ref{PEsublattice2}) the parallel execution time was directly compared with $1/N_p$ times the serial execution time of a system whose total system size is the same as   in the parallel simulation. If the serial KMC  calculation time per event is independent of   system size, then there should be no difference between the two definitions.  However, in general  this may not be the case.  For example, in KMC simulations in which the rates for all events are stored separately,  the calculation time per event will increase as $M^{1/2}$ (where $M = L_x L_y$ is the system size) using the Maksym algorithm\cite{maksym}  and as  log($M$) using a  binary tree algorithm.\cite{blue}  In  this case, the parallel efficiency calculated using Eq. \ref{PEsublattice2} may be significantly higher than that obtained using Eq. \ref{PEsublattice}, and may even be larger than $1$, since the division of a system into smaller subsystems may reduce the calculation time per event per processor.

Since in the   models studied here we have used lists for each type of event,  rather than the ``partial-sum" algorithms described above, we would expect the serial calculation time per event to be   independent of  system size, and thus the two definitions of parallel efficiency should  be equivalent. To test if this is   the case, we have calculated the serial simulation time per event for the fractal model for $D/F = 10^3$ and $D/F = 10^5$ for a variety of system sizes ranging from $L = 64$ to $L = 2048$.  Somewhat surprisingly, we found that the serial calculation time per event increases slowly with increasing processor size.  In particular,  an increase of approximately 50\%  in  the calculation time per event was obtained when going from a system of size $L = 64$ to  $L = 2048$.    We believe that this is most likely due to  memory or ``cache" effects in our simulations.  This increase in the serial calculation time per event with increasing system size indicates that the  calculated  parallel efficiencies  shown in Fig. \ref{Figpenp} would  actually be somewhat  larger if the more direct definition of parallel efficiency (Eq. \ref{PEsublattice2}) were used.     
However, since for large $N_p$ this requires serial simulations of very large systems (e.g. $256,000 \times 256,000$ for $N_p = 100$), the first definition (Eq. \ref{PEsublattice}) was used. 

Finally, we return to the general question of the   applicability and validity of the SL algorithm.  In general, we expect that   for  a wide class of non-equilibrium processes there exists a clearly defined diffusion length $l_D$, which may or may not vary slowly with time.  For these processes we expect that  finite-size effects will not occur as long as the sublattice size is larger than $l_D$.  Furthermore, as long as this length scale is not too large compared to the  desired  system size,  parallel simulations using the SL algorithm will be advantageous.  As our results indicate, parallel KMC simulations using the synchronous sublattice algorithm are in general likely to be significantly faster than  either the conservative asynchronous algorithm or the synchronous relaxation algorithm.  As a result, we expect that the synchronous sublattice algorithm   may be particularly useful as a means to carry out a variety of   parallel non-equilibrium simulations.

\begin{acknowledgments}

This research was supported by the NSF through
Grant No. DMR-0219328.  We would also like to acknowledge grants of computer time from the Ohio Supercomputer Center (Grant no. PJS0245) and the  Pittsburgh Supercomputer
Center (Grant no. DMR030007JP). 

\end{acknowledgments}

\end{document}